\documentclass[pra,11pt]{revtex4-1}

\usepackage{amsmath}    % need for subequations
\usepackage{amssymb}
\usepackage{graphicx}   % need for figures
\usepackage{verbatim}   % useful for program listings
\usepackage{color}      % use if color is used in text
\usepackage{epsfig}
\usepackage{subfigure}  % use for side-by-side figures
\usepackage{hyperref}   % use for hypertext links, including those to external documents and URLs
\usepackage{accents}
\newcommand\thickbar[1]{\accentset{\rule{.6em}{.8pt}}{#1}}

\newcommand{\bea}{\begin{eqnarray}}
\newcommand{\eea}{\end{eqnarray}}
\newcommand{\ket}[1]{\left|{#1}\right\rangle}
\newcommand{\bra}[1]{\left\langle{#1}\right|}

\newcommand{\inner}[2]{\left\langle{#1}|{#2}\right\rangle}

\begin{document}
\title{\textbf{$q$-deformed quadrature operator and optical tomogram}}
\author{{M. P. Jayakrishnan$^{1}$, Sanjib Dey$^{2,3}$, Mir Faizal$^{4,5}$ and C. Sudheesh$^{6}$} \\ \footnotesize{$^{1}$School  of  Physics, Indian Institute of Science Education and Research Thiruvananthapuram, India 695 016\\ $^{2}$Institut des Hautes \'Etudes Scientifiques, Bures-sur-Yvette 91440, France \\ $^{3}$Institut Henri Poincar\'e, Paris 75005, France \\ $^{4}$Irving K. Barber School of Arts and Sciences, University of British Columbia-Okanagan Kelowna, British Columbia V1V 1V7, Canada \\ $^{5}$Department of Physics and Astronomy, University of Lethbridge, Lethbridge, Alberta T1K 3M4, Canada \\ $^{6}$Department of Physics, Indian Institute of Space Science
and Technology, Thiruvananthapuram, India  695 547\\ Email: jayakrishnan00213@iisertvm.ac.in, sdey@ihes.fr, mirfaizalmir@googlemail.com, sudheesh@iist.ac.in}}

%\date{24 October 2014}

%\pacs{03.65.Wj}{Quantum tomography}
%\pacs{42.50.Dv}{Quantum state engineering and measurements}

\begin{abstract}

In this paper, we define the homodyne $q$-deformed  quadrature operator and find its eigenstates in terms of the deformed Fock states. 
We find the   quadrature representation of  $q$-deformed Fock states in the process.
Furthermore, we calculate  the explicit analytical expression for the optical tomogram of the $q$-deformed coherent states.

 \end{abstract}

\maketitle
\section{Introduction}\label{sec1}

\addtolength{\voffset}{-0.8cm} %% Adds extra space in the first page of the article
\addtolength{\footskip}{0.8cm} %% Adds extra space in the footer
The general principle behind quantum tomography is that instead of extracting a particular property of a quantum state (e.g. quantum  entanglement), it aims to extract all possible information about the state that are contained in the density operator. Quantum tomography characterizes the complete quantum state of a particle or particles through a series of measurements in different quantum systems described by identical density matrices, much like its classical counterpart, which aims at reconstructing three-dimensional images via a series of two-dimensional projections along various directions. In optical phase space, the position and momentum of a quantum particle are determined by the quadratures. By measuring one of the quadratures of a large number of identical quantum states, one obtains a probability density corresponding to that particular quadrature, which characterizes the particle's quantum state. Thus, the quantum tomogram is defined as the probability that the system is in the eigenstate of the quadrature operator\cite{barnet}.

Quantum tomography is often used for analyzing optical signals, including measuring the signal gain and loss of optical devices \cite{Ariano}, as well as in quantum computing and quantum information theory to reliably determine the actual states of the qubits \cite{revmod}. As for instance, one can imagine a situation in which a person Bob prepares some quantum states and then sends the states to Alice to look at. Not being confident with Bob's description of the states, Alice may wish to do quantum tomography to classify the states herself. Balanced homodyne detection provides an experimental technique to study the quantum tomogram \cite{revmod,homodyne}, which is a probability distribution of homodyne quadrature depending on an extra parameter of local oscillator phase $\theta$. When $\theta$ is varied over a whole cycle, it becomes the tomogram and, thus, tomogram contains complete information about the system. Quasi-probabilistic distributions describing the state of the system can be reconstructed from the tomogram via transformations like inverse Radon transformations\cite{radon}. In \cite{tomo1}, the authors deal with the tomography of photon-added coherent states, even and odd coherent states, thermal states etc. The tomogram of coherent states as well as the evolution of tomogram of a state in a nonlinear medium was studied in \cite{tomogram}, which essentially demonstrated the signatures of revivals, fractional revivals and decoherence effects (both amplitude decay and phase damping) in the tomogram. Recently, the signatures of entanglement was observed theoretically   in the optical tomogram 
of the quantum state  without reconstructing the density matrix of the system \cite{entanglement}. 
A detailed discussion on the  formulation of quantum mechanics using tomographic probabilities has been reported in \cite{ibort}.

On the other hand, $q$-deformed oscillator algebras have been very famous in various subjects during last few decades, which were introduced through a series of articles \cite{beiden,macfar,sun,kulish}. There are mainly two kinds of deformed algebras, namely, maths type \cite{math,arik,sanjib} and physics type \cite{beiden,macfar,rogers}.  Algebras of both types have been utilized to construct $q$-deformed bosons having applications in many different contexts, in particular, in the construction of coherent states \cite{beiden,eremin,sanjib}, cat states \cite{Mancini,Dey}, photon-added coherent states \cite{Naderi,Dey_Hussin}, atom laser \cite{atom}, nonideal laser \cite{nonli}, etc. Besides, they are frequently used on the study of quantum gravity \cite{Major}, string theory \cite{Aganagic}, non-Hermitian Hamiltonian systems \cite{bagchi,sanjib,Dey}, etc. The principal motivation of the present article is to study a method of quantum tomography for $q$-deformed coherent states by considering the maths type deformed canonical variables studied in \cite{bagchi,sanjib}. We also introduce the $q$-deformed homodyne quadrature related to the above mentioned deformed algebra, which is one of the principal requirements for the study of quantum tomography.

Our paper is organized as follows: In Sec. \ref{sec2}, we define the $q$-deformed homodyne quadrature operator. The eigenstates of the deformed quadrature have been found analytically   in Sec. \ref{sec3}. In the process, we also find the quadrature representation of the deformed Fock states.  In Sec. \ref{sec4}, we provide a short review of the  optical tomography followed by the tomography of $q$-deformed coherent states. Finally, our conclusions are stated in Sec. \ref{sec5}.
%%%%%%%%%%%%%%%%%%%%%%%%%%%%%%%%%%%%%%%%%%%%%%%%%%%%%%%%%%%%%%%%%%%%%%%%%%%%%%%%%
% Section 2
%%%%%%%%%%%%%%%%%%%%%%%%%%%%%%%%%%%%%%%%%%%%%%%%%%%%%%%%%%%%%%%%%%%%%%%%%%%%%%%%%
\section{$q$-deformed quadrature operator}\label{sec2}
Let us commence with a brief discussion of a $q$-deformed oscillator algebra introduced in \cite{arik,bagchi,sanjib}
\addtolength{\voffset}{0.8cm} %% Adds extra space
\addtolength{\footskip}{-0.8cm} %% Adds extra space in the footer
%\lhead{$q$-deformed quadrature operator and optical tomogram}
%\chead{}
%\rhead{}
\begin{equation}\label{1}
AA^\dag-q^2 A^\dag A=1, \qquad |q|<1,
\end{equation} 
which is often known as the math type $q$-deformation in the literature. As obviously, in the limit $q\rightarrow 1$, the $q$-deformed algebra (\ref{1}) reduces to the standard canonical commutation relation $[a,a^\dagger]=1$. The deformed algebra  has been used before in describing plenty of physical phenomena \cite{sanjib,Dey,Dey_Hussin}. Moreover, a concrete Hermitian representation of the corresponding algebra was derived in \cite{sanjib} by utilizing the Rogers-Sz\"ego polynomial \cite{rogers} with the operators $A,A^\dagger$ being bounded on the region of unit circle. The deformed algebra given in Eq. (\ref{1}) can be defined on the $q$-deformed Fock space forming a complete orthonormal basis provided that there exists a deformed number operator $[n]$ of the form
\begin{equation}
[n]=\frac{1-q^{2n}}{1-q^2},
\end{equation}    
such that the action of the annihilation and creation operators on the Fock states $\ket{n}_q$ are given by
\begin{eqnarray}
A\ket{n}_{q} &=& \sqrt{[n]} \ket{n-1}_{q}, \qquad A\ket{0}_{q}=0, \label{2}\\
A^\dag \ket{n}_{q} &=& \sqrt{[n+1]}\ket{n+1}_{q}. \label{3}
\end{eqnarray}
In the limit $q\rightarrow 1$, the deformed Fock state $\ket{n}_q$ reduces to the Fock state, $\ket{n}$, which is an eigenstate of the operator $a^{\dagger}a$ with eigenvalue $n$. It is  possible to define a set of canonical variables $X,P$ in terms of the $q$-deformed oscillator algebra generators
\begin{equation}\label{5}
X=\alpha (A^\dag+A), \quad P=i\beta (A^\dag -A),
\end{equation}
with $\alpha =\beta =\dfrac{\sqrt{1+q^2}}{2}$ satisfying the deformed commutation relation \cite{sanjib}
\begin{equation}\label{6}
[X,P]=i\Big[1+\dfrac{q^2-1}{q^2+1}(X^2+P^2)\Big].
\end{equation}
Let us now define the homodyne $q$-deformed quadrature operator 
\begin{equation}\label{7}
\hat{X}_{\theta}=\dfrac{\sqrt{1+q^2}}{2}(Ae^{-i\theta}+A^\dag e^{i\theta}),
\end{equation}
with $\theta $ being the phase of the local oscillator associated with the homodyne detection setup such that $ 0\leq \theta \leq 2\pi $. Clearly at $\theta =0$ and $\pi/2$, one obtains the dimensionless canonical observables $X$ and $P$, respectively.  The  definition given in Eq. (\ref{7}) is consistent with the homodyne detection theory \cite{Braun, Vogel, Banas}. In the limit $q\rightarrow 1$, the quadrature operator $\hat{X}_{\theta}$ reduces to the quadrature operator, 
\begin{equation}\label{qfock}
\hat{x}_{\theta}=\dfrac{1}{\sqrt{2}}(\hat{a}e^{-i\theta}+\hat{a}^\dag e^{i\theta}),
\end{equation}
in the non-deformed algebra $[a,a^\dagger]=1$.

%%%%%%%%%%%%%%%%%%%%%%%%%%%%%%%%%%%%%%%%%%%%%%%%%%%%%%%%%%%%%%%%%%%%%%%%%%%%%%%%%
%  Section 3
%%%%%%%%%%%%%%%%%%%%%%%%%%%%%%%%%%%%%%%%%%%%%%%%%%%%%%%%%%%%%%%%%%%%%%%%%%%%%%%%%
\section{Eigenstates of the $q$-deformed quadrature operator}\label{sec3}  
This section contains the explicit calculation of the eigenstate of the $q$-deformed quadrature operator $\hat{X}_{\theta}$:
\begin{equation}\label{8}
\hat{X}_{\theta}\ket{X_{\theta}}_q=X_{\theta}\ket{X_{\theta}}_q,
\end{equation}
with $X_{\theta}$ being the eigenvalue. By using Eqs.  (\ref{2}), (\ref{3}) and (\ref{7}), we obtain
\begin{eqnarray}
{}_q\bra{n}\hat{X}_{\theta}\ket{X_{\theta}}_q =X_{\theta}\thickbar{\Psi}_{n_{q}}(X_{\theta}) \label{9}&=& \dfrac{\sqrt{1+q^2}}{2}{}_q\bra{n}(Ae^{-i\theta}+A^\dag e^{i\theta})\ket{X_{\theta}}_q \\
&=& \dfrac{\sqrt{1+q^2}}{2}\Big(\sqrt{[n+1]}e^{-i\theta}{}_q\langle n+1|X_\theta\rangle_q+\sqrt{[n]}e^{i\theta}{}_q\langle n-1|X_\theta\rangle_q\Big) \\
&=& \dfrac{\sqrt{1+q^2}}{2}\Big(\sqrt{[n+1]}e^{-i\theta}\thickbar{\Psi}_{n+1_{q}}(X_{\theta})+\sqrt{[n]}e^{i\theta}\thickbar{\Psi}_{n-1_{q}}(X_{\theta})\Big), \label{10}
\end{eqnarray}
where we denote ${}_q\langle n|X_\theta\rangle_q$, ${}_q\langle n+1|X_\theta\rangle_q$ and ${}_q\langle n-1|X_\theta\rangle_q$ by $\thickbar{\Psi}_{n_{q}}(X_{\theta})$, $\thickbar{\Psi}_{n+1_{q}}(X_{\theta})$ and $\thickbar{\Psi}_{n-1_{q}}(X_{\theta})$, respectively. 
The complex conjugate of $\thickbar{\Psi}_{n_{q}}(X_{\theta})$ gives the quadrature representation of the deformed Fock state $\ket{n}_q$:
\begin{equation}
\Psi_{n_{q}}(X_{\theta})={}_q\langle X_\theta |n\rangle_q.
\label{qrepresentation}
\end{equation} 
 When $\theta=0$, the wave function $\Psi_{n_{q}}(X_{\theta=0})$  corresponds to the position representation of the deformed Fock state.  Henceforth, we use $\Psi_{n_{q}}(X_{\theta})$ in the calculation instead of $\thickbar{\Psi}_{n_{q}}(X_{\theta})$ because the former is directly  the quadrature representation of the deformed Fock state to obtain.  
After taking the complex conjugate of the  Eq. (\ref{10}) and rearranging the terms in it,   we get  a three term recurrence relation for $\Psi_{n_{q}}(X_{\theta})$:
\begin{equation}\label{11}
{\Psi}_{n+1_{q}}(X_{\theta})=\frac{e^{-i\theta}}{\sqrt{[n+1]}}\left[\frac{2}{\sqrt{1+q^2}}X_\theta
{\Psi}_{n_{q}}(X_{\theta})-\sqrt{[n]}\,{\Psi}_{n-1_{q}}(X_{\theta})e^{-i\theta}\right].
\end{equation}
First few terms of which are
\begin{alignat}{1}
{\Psi}_{1_q} &= \frac{e^{-i\theta}}{\sqrt{[1]}}\frac{2 X_\theta}{\sqrt{1+q^2}}\,{\Psi}_{0_q}(X_\theta) \\
{\Psi}_{2_q} &= \frac{e^{-2i\theta}}{\sqrt{[2]}}\left[\frac{2 X_\theta}{\sqrt{1+q^2}}\left(\frac{2 X_\theta}{\sqrt{[1](1+q^2)}}\right)-\sqrt{[1]}\right]\,{\Psi}_{0_q}(X_\theta) \\
{\Psi}_{3_q} &= \frac{e^{-3i\theta}}{\sqrt{[3]}}\left[\frac{2 X_\theta}{\sqrt{1+q^2}}\frac{1}{\sqrt{[2]}}\left(\frac{2 X_\theta}{\sqrt{1+q^2}}\frac{2 X_\theta}{\sqrt{[1](1+q^2)}}-\sqrt{[1]}\right)-\sqrt{[2]}\frac{2 X_\theta}{\sqrt{[1](1+q^2)}}\right] {\Psi}_{0_q}(X_\theta).\label{psi3q}
\end{alignat}
Using Eqs. (\ref{11}-\ref{psi3q}),  we find the analytical expression for $q$-deformed Fock state $\ket{n}_q$  in the quadrature basis  as
\begin{equation}\label{12}
\Psi_{n_{q}}(X_{\theta})=\textit{J}_{n_{q}}(X_{\theta}) e^{-in\theta}\Psi_{0_{q}}(X_{\theta}).
\end{equation}
Here,  we introduce the   new polynomial $\textit{J}_{n_{q}}(X_{\theta})$ which is  defined by the following recurrence relation
\begin{equation}\label{13}
\textit{J}_{n+1_{q}}(X_{\theta})=\frac{1}{\sqrt{[n+1]}}\left[\dfrac{2X_{\theta}}{\sqrt{1+q^2}}J_{n_{q}}(X_{\theta})-\sqrt{[n]}J_{n-1_{q}}(X_{\theta})\right],
\end{equation}
with $J_{0_{q}}(X_{\theta})=1$ and $J_{1_{q}}(X_{\theta})=2X_{\theta}/\sqrt{[1](1+q^2)}$. In order to check the consistency, we take the limit $q\rightarrow 1$ and, indeed in the limiting condition the wavefunction $\Psi_{n_{q}}(X_{\theta})$ given in Eq.  (\ref{12}) reduces to the quadrature representation of the 
Fock state $\ket{n}$:
\begin{equation}
\Psi_{n_{q\rightarrow 1}}(X_{\theta}\rightarrow x_\theta)= \dfrac{H_{n}(x_{\theta})}{\pi^{1/4}\,2^{n/2}\sqrt{n!}}\,e^{-in\theta}
e^{-x_{\theta}^2/2},
\end{equation}
with $H_{n}(x_{\theta})$ being the Hermite polynomial of order $n$ and identifying
\begin{equation}
\Psi_{0_{q\rightarrow 1}}(X_{\theta}\rightarrow x_\theta)=\frac{e^{-x_{\theta}^2/2}}{\pi^{1/4}}.
\end{equation}
Correspondingly, the recurrence relation in Eq.  (\ref{13})  merges with the recurrence relation of the Hermite polynomials
\begin{equation}\label{15}
H_{n+1}(x_{\theta})=2x_{\theta}H_{n}(x_{\theta})-2nH_{n-1}(x_{\theta}).
\end{equation}

Next, we calculate the eigenstate of the $q$-deformed quadrature operator. % using the quadrature representation 
%of the deformed Fock state, $\Psi_{n_{q}}(X_{\theta})$,  obtained in Eq. (\ref{12}).  
 By using  Eqs. (\ref{qrepresentation}) and (\ref{12}),   we derive  the explicit expression for the eigenstates of $q$-deformed quadrature operator  $\hat{X}_{\theta}$  as follows: 
\begin{equation}\label{16}
\ket{X_{\theta}}_q=\sum_{n=0}^{\infty}\ket{n}_{q}{}_q\langle n| X_{\theta}\rangle_q=\thickbar{\Psi}_{0_{q}}(X_{\theta})\sum_{n=0}^{\infty}\textit{J}_{n_{q}}(X_{\theta}) e^{in\theta}\ket{n}_{q},
\end{equation}
with $\Psi_{0_{q}}(X_{\theta})$ being the ground state wavefunction in the deformed quadrature basis such that
\begin{equation}\label{17}
{}_q\inner{X^\prime_{\theta}}{X_{\theta}}_q=\delta(X_{\theta}-X^{\prime}_{\theta})=\thickbar{\Psi}_{0_{q}}(X_{\theta})\,{\Psi}_{0_{q}}(X^{\prime}_{\theta}) \sum_{n=0}^{\infty} \textit{J}_{n_{q}}(X_{\theta})\textit{J}_{n_{q}}(X^{\prime}_{\theta}).
\end{equation}
When we take the limit $q\rightarrow 1$ in the expression given in Eq. (\ref{16}), we get the eigenstates of the quadrature operator $\hat{x}_{\theta}$ \cite{barnet}:
\begin{equation}\label{20}
\ket{x_{\theta}}=\dfrac{1}{\pi^{1/4}}\sum_{n=0}^{\infty}\dfrac{e^{in\theta}}{\sqrt{n!}}\dfrac{1}{2^{n/2}}H_{n}(x_{\theta})e^{-x_{\theta}^2/2}\ket{n}.
\end{equation}
In the following section, we use the eigenstates  $\ket{X_\theta}_q$ obtained in Eq. (\ref{16}) to calculate the optical tomogram of the $q$-deformed coherent state. 
%%%%%%%%%%%%%%%%%%%%%%%%%%%%%%%%%%%%%%%%%%%%%%%%%%%%%%%%%%%%%%%%%%%%%%%%%%%%%%%%%
%  Section 4
%%%%%%%%%%%%%%%%%%%%%%%%%%%%%%%%%%%%%%%%%%%%%%%%%%%%%%%%%%%%%%%%%%%%%%%%%%%%%%%%%
\section{$q$-deformed optical tomography}\label{sec4} 
In order to find the  optical tomogram of the $q$-deformed coherent states, let us first briefly recall the notions of the optical tomography.  For a state of the system represented by the density matrix $\hat{\rho}$, the optical tomogram $\omega(X_{\theta},\theta)$ is given by the expression
\begin{equation}\label{19}
\omega(X_{\theta},\theta)=\langle X_{\theta}|\hat{\rho}| X_{\theta}\rangle,
\end{equation}
with the normalization condition
\begin{equation}\label{21}
\int \omega(X_{\theta},\theta)\,dX_{\theta}=1,
\end{equation}
where $\ket{X_{\theta}}$ is the eigenstate of  homodyne quadrature operator  $\hat{X}_{\theta}$ with eigenvalue $X_{\theta}$. Thus, the tomogram of a pure state represented by the density matrix $\hat{\rho}=\ket{\Phi}\bra{\Phi}$ is given by the expression $\omega (X_{\theta},\theta)=\mid\inner{X_{\theta}}{\Phi}\mid^2$\cite{tomogram,tomo1,tomo2}. Here, we are interested to compute the tomogram of the $q$-deformed coherent states \cite{arik}
\begin{equation}\label{22}
\ket{\Phi}_q=\frac{1}{\sqrt{E_{q}(\mid \alpha \mid^2))}}\sum_{n=0}^{\infty}\dfrac{\alpha^n}{\sqrt{[n]!}}\ket{n}_{q},\quad [n]!=\displaystyle\prod_{k=1}^{n}[k],\quad [0]!=1,
\end{equation}
where $\alpha \in \mathbb{C}$ and
\begin{equation}
E_{q}(\mid \alpha \mid^2)=\sum_{n=0}^{\infty}\dfrac{|\alpha |^{2n}}{[n]!}.
\end{equation}
Given the eigenstates (\ref{16}) of the $q$-deformed quadrature operator $\hat{X}_{\theta}$, we find the tomogram of the above $q$-deformed coherent states $\ket{\Phi}_q$ as follows
\begin{equation}\label{23}
\omega (X_{\theta},\theta)=\Bigg|\sum_{n=0}^{\infty}\dfrac{\alpha^n \textit{J}_{n_{q}}(X_{\theta}) e^{-in\theta}\Psi_{0_{q}}(X_{\theta})}{\sqrt{E_{q}(\mid \alpha \mid^2)}\sqrt{[n]!}}\Bigg|^2.
\end{equation}
In the limit $q\rightarrow 1$, the above tomogram   $\omega (X_{\theta},\theta)$   become 
the tomogram of the Glauber coherent states $\ket{\alpha}$ \cite{barnet}
\begin{equation}\label{cstomogram}
\omega (x_{\theta},\theta)=\frac{1}{\pi^{1/4}}\,\exp\left( -\frac{x_\theta^2}{2}-\frac{|\alpha|^2}{2}-\frac{\alpha^2\,e^{-i2\theta}}{2}+\sqrt{2}\alpha\,x_\theta\, e^{-i\theta}\right),
\end{equation}
which corroborate   the   expression given in Eq. (\ref{23}) for the tomogram of $q$-deformed coherent state.
%%%%%%%%%%%%%%%%%%%%%%%%%%%%%%%%%%%%%%%%%%%%%%%%%%%%%%%%%%%%%%%%%%%%%%%%%%%%%%%%%
%  Section 5
%%%%%%%%%%%%%%%%%%%%%%%%%%%%%%%%%%%%%%%%%%%%%%%%%%%%%%%%%%%%%%%%%%%%%%%%%%%%%%%%%
\section{Conclusions}\label{sec5} 
We defined a $q$-deformed quadrature operator  compatible with the homodyne detection technique and  found its eigenstates  in terms of a new $q$-deformed polynomial.  The eigenstates of the quadrature operator obtained in this paper are very important  because  they enable us to find the quadrature representation of any $q$-deformed state.  These eigenstates are also required to find theoretically the optical tomogram of the quantum states and compare it with experimentally obtained tomogram.  We found the quadrature representation of the deformed Fock states  and confirmed it by checking the limiting case. These quadrature representations can be used to find easily the quasi-probability distributions of deformed quantum states.   Finally, the $q$-deformation of the quantum tomography has been found by utilizing the expression for the eigenstates of the $q$-deformed quadrature operator.

\vspace{0.5cm} \noindent \textbf{\large{Acknowledgements:}} MPJ is supported by a INSPIRE Fellowship by Department of Science and Technology (DST), Government of India, and SD is supported by a CARMIN Postdoctoral Fellowship by the Institut des Hautes \'Etudes Scientifiques and Institut Henri Poincar\'e.
%%%%%%%%%%%%%%%%%%%%%%%%%%%%%%%%%%%%%%%%%%%%%%%%%%%%%%%%%%%%%%%%%%%%%%%%%%%%%%%%%
%  References
%%%%%%%%%%%%%%%%%%%%%%%%%%%%%%%%%%%%%%%%%%%%%%%%%%%%%%%%%%%%%%%%%%%%%%%%%%%%%%%%%

%\bibliographystyle{unsrt} 
%\bibliography{Ref.bib}
%\input{Reference.tex}

\end{document}